# Seeds and Sequences of Element Abundances in Solar Energetic Particle Events

**Donald V. Reames** (https://orcid.org/0000-0001-9048-822X )
Institute for Physical Science and Technology,
University of Maryland, College Park, MD, USA
dvreames@gmail.com

**Abstract** Solar energetic particles (SEPs) in the small "impulsive" events, primarily accelerated during magnetic reconnection in solar jets, have strong enhancements of the abundances of increasingly heavy elements. In contrast, the shock acceleration of ambient coronal plasma in most large "gradual" SEP events produces flat or decreasing abundances vs. element mass-to-charge ratios $A/Q$. However, heavy-ion enhancements in the largest gradual SEP events can occur in two ways: (1) strong streaming of protons away from the shock amplifies Alfvén waves that preferentially scatter and retard protons near the shock while increasingly-heavy ions can leak out, and (2) strong shock waves reaccelerate SEPs fed from persistent impulsive SEP events streaming from some active regions, with their pre-enhanced heavy ions becoming dominant. Power-law fits of abundance enhancements vs. $A/Q$ can distinguish the latter events by the presence of both impulsive and coronal seed components and the best-fit charges $Q$ define characteristic source temperatures. Ironically, ions with high observed charges, e.g. $Q_{Fe} \approx 20$, are yet another signature of impulsive seed ions that are routinely stripped after initial acceleration. Intense impulsively-seeded events can occur in sequences fed from a single persistent active-region as it rotates across the disk of the Sun. Three week-long event sequences, each producing two or three very large events, occur early in the strong solar-cycle 23 (1997 – 2008). The weak solar cycle 24, produces only one impulsively-seeded event sequence – perhaps a dearth of both impulsive seeds and sufficiently strong shocks. Solar cycle 25 has produced an unusual active period of short strong impulsive events. In contrast, there are other active regions where large events alternate SEPs with and without impulsively-seeded sources. We also find that events with moderate Alfvén-wave trapping near the shock can release ions slowly or rapidly as a function of $A/Q$. This $A/Q$-dependent trapping acts almost as a magnetic spectrometer that separates elements in space and time.

# 1 Introduction

It has taken many years to confirm two primary processes for the acceleration of solar energetic particles (SEPs). They are accelerated (1) as "impulsive" SEP events by magnetic reconnection in (open) solar jets and in (closed) flares, and (2) as "gradual" SEP events by fast collisionless shock waves driven out from the Sun by coronal mass ejections (CMEs), the latter process producing the largest events. However, these two major processes are not exclusive. Shock waves can reaccelerate ions supplied by impulsive SEPs, especially when fed by persistent streams of impulsive events, as well as ions from the ambient corona. How can the pattern of element abundances distinguish these different seed populations? How important are events with reaccelerated impulsive seeds? How long can streaming sources of impulsive seed particles persist? Do they come and go or can they seed several large events from a single active region? How does this depend upon the strength of the solar cycle? In this article we use measurements of element abundances in SEP events to probe the persistence of the sources of those elements from event to event and their availability from cycle to cycle and to examine other physical processes that control abundance variations.

The earliest evidence of energetic particles from the Sun was reported by Forbush in 1946 [1] when his measurements of ground-level muons from GeV atmospheric reactions showed intensity peaks preceding the galactic-cosmic-ray (GCR) decreases that bear his name. Much later we would find that, ironically, these ground-level enhancements (GLEs) were produced by GeV protons accelerated by the shock wave driven by the same fast, wide CMEs that screened out GCRs to produce the Forbush decreases.

Elements, C, O, and up to S were first distinguished by Fichtel and Guss [2] counting the hairy delta-rays tracks of scattered electrons along the ion tracks in nuclear emulsion detectors flown on sounding rockets from Ft. Churchill, Manitoba. Abundances were extended up to Fe by the same technique [3]. Satellite-borne instruments improved both the coverage and resolution [4, 5], abundances were found to vary little with energy, and comparison with solar abundances Meyer [6] found that SEPs reflected stable coronal abundances underlying positive and negative power-law dependence on their mass-to-charge ratio $A/Q$ as seen by Breneman and Stone [7]. Reference coronal abundances of major elements became relatively well defined from averages over many large SEP events [8 – 10]. Those reference coronal abundances from SEPs differ from solar photospheric abundances as a function of the first ionization potential (FIP) of the element. Low-FIP (<10 eV) elements are ionized in the photosphere and become enhanced in transiting the chromosphere to the corona, probably because of the ponderomotive forces of Alfvén waves [11] that cannot affect the high-FIP elements that are initially neutral atoms. All atoms become ionized in the ~1-MK corona. The SEP corona differs from that producing the solar wind, with its higher boundary between high and low FIP ions [10, 12-17].

The discovery of $^3$He-rich SEP events, some with more $^3$He than $^4$He [18] introduced new physics for small events that were found to be associated with the frequent type III radio bursts generated by streaming solar electrons [19, 20]. Early observers thought that the $^3$He might be produced by nuclear fragmentation as it is in GCRs, but the lack of $^2$H and especially the lack of Li, Be, and B, found to be < 2 ×10$^{-4}$ of O [5, 21] showed otherwise, even though γ-ray lines show significant fragmentation in flares – but those ions are magnetically trapped. Many wave modes were proposed for resonantly heating $^3$He (see references in [10] or [16]) before Temerin and Roth [22] suggested that it was electromagnetic ion-cyclotron (EMIC) waves from those streaming type-III electrons that resonantly accelerated the $^3$He mirroring in converging fields. In a new simulation by Fitzmaurice et al. [23], waves produced by H and $^4$He streaming from the acceleration site resonantly heat $^3$He and scatter it back into the acceleration region and Jain et al. [24] have used 2D hybrid-kinetic simulations to study the power-law abundance enhancements in *A/Q*.

These impulsive SEP events were also found to be Fe-rich [25], an enhancement that would be extended with improving observations [26, 27], eventually to elements as heavy as Pb [28 – 32], as the 3.6 power of *A/Q*, on average, with *Q*-values determined at a temperature of ~2.5 MK [32, 33]. Early observations found $^3$He-rich events to be unrelated to CMEs [34] but more-sensitive coronagraphs showed associations with narrow CMEs from solar jets [35]; jets had been recognized as sources of type III bursts [36] and now also as the sources of impulsive SEP events [37 – 43]. Theoretically, particle-in-cell simulations by Drake et al. [44] (see also Kramoliš et al. [45]) found that the power-law enhancements in *A/Q* were related to ion gyro-radii and the width distribution of collapsing islands of magnetic reconnection.

Actually, the earliest evidence of the two different sources of SEPs we now call impulsive and gradual events, was in the radio observations of type III bursts of solar streaming electrons and type II bursts from electrons accelerated by shocks [46]. The source of shocks and type II bursts became clear when 96% of large SEP events were found to be associated with CME-driven shock waves by Kahler et al. [47], but a decade later it was still necessary to emphasize the importance of shock acceleration in "The Solar Flare Myth" by Gosling [48, 49]. Shock acceleration of SEPs has now been established and studied widely in many different ways [50-77].

The complexity of seed particles at shocks emerged when Mason et al. [78] found enhanced $^3$He in a large gradual SEP event. Other encroachments of impulsive suprathermal ions were soon found [79 – 82] and it was thought that these suprathermal ions might be favored when the magnetic field was quasi-perpendicular to the shock normal where these initially-faster impulsive-SEP ions could overtake the shock from downstream [83].

However, evidence of persistent sources of impulsive seed ions has been available for many years. Figure 1 replots Fe, O, and electron intensities during a three-day period in April 1980 originally shown by Reames [84]. The multiple small impulsive events are difficult to distinguish in Fe and O, but the average Fe/O = 4.2 ± 1.4 for the period, 32 times the coronal abundance. This ratio is unusually high, but these persistent impulsive streams are not so unusual. It is easy to imagine that such a stream of many impulsive events could provide a rich target for a large, fast CME-driven shock wave from the same active region to produce a heavy-element-dominated gradual SEP event.

In fact, many of the $^3$He-rich "events" we encounter may not be single events, but complicated streams with poorly-resolved multiple sources [37]. Many observers have reported evidence for persistent streams of impulsive suprathermal ions [80, 85 – 88]. Parashiv and Donea [89] have identified "coronal geysers" that can produce a dozen recurrent jets and type III bursts daily in active regions.

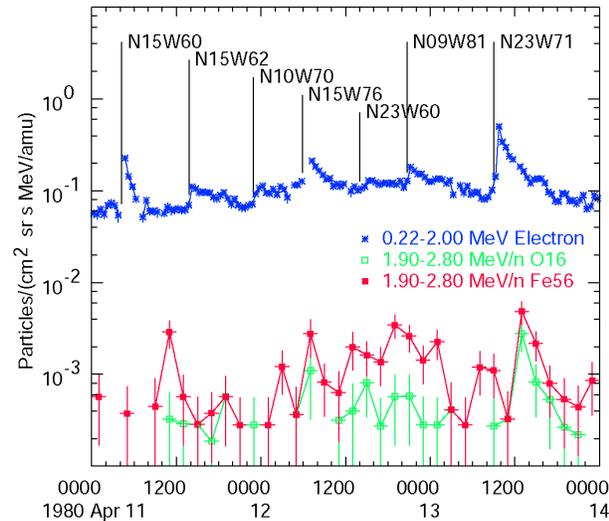

**Fig. 1** ISEE 3 measurements of intensities of electrons, O, and Fe, at the listed energies, are shown vs. time for a four-day period in April 1980. Electron increases suggest many small SEP events that continue to contribute a blur of Fe-rich ions from a passing active region. Such streams have been seen many years, and can be targets with seed ions for a large shock wave to produce a heavy-element-rich SEP3 event. Of course, we see these events here because they have *not* been disrupted by a shock. Such streams of impulsive ions are magnetically connected to Earth for days, but they may persist above an active region much longer.

In an effort to account for this newfound variety of seed populations and acceleration sites, Reames [90] proposed four categories of SEP events:

(1) SEP1 "pure" impulsive events from reconnection in jets with no fast CMEs.

(2) SEP2 impulsive events with reacceleration by a local fast (>500 km s$^{-1}$) CME-driven shock from the same jet.

(3) SEP3 gradual events with fast, wide CMEs where high-$Z$ ions are dominated by seed particles from collected multi-jet streams of SEP1 ions like Fig 1.

(4) SEP4 gradual event with fast, wide CMEs dominated at all $Z$ by ions from ambient coronal seed particles.

Why should our categories depend so much upon reaccelerated Fe-rich populations rather than being defined by the uniquely-impulsive $^3$He ions themselves? Unfortunately $^3$He/$^4$He varies greatly in impulsive events and even varies orders of magnitude from energy to energy within a single event [31]. Elevated Fe/O (or a power-law in $A/Q$) is a much more stable indicator of impulsive events [32]. While Fe/O can also increase from differential transport in very large SEP4 events, we will find that that behavior is rather clear and distinguishable.

Section 2 reviews the process of fitting abundance enhancements, Sect. 3 shows sequences of SEP3 events, and Sect. 4 studies contrasting sequences with SEP4 events, followed by Discussion and Summary.

Most SEP abundances in this work were measured by the low-energy matrix telescope (LEMT) on the *Wind* spacecraft [91, 92] (see also Chapt. 7 in [10]) and are available from https://omniweb.gsfc.nasa.gov/ . CME speeds were obtained from the *Large Angle and Spectrometric Coronagraph* (LASCO) on the *Solar and Heliospheric Observatory* (SOHO) at https://cdaw.gsfc.nasa.gov/CME_list/.

## 2 Determining *A/Q*, Power-Law Fits, and Temperatures

Energetic-particle abundance enhancements are found using the measured abundance of each element in an event or time interval, relative to the abundance of O at the same energy-per-nucleon, and dividing by the corresponding coronal reference abundance, also relative to O. In order to plot these enhancements vs. *A/Q*, we need to know the average ionization state of each element or the temperature that determined *Q* during acceleration.

There is a history of direct charge-state measurements of SEPs. Luhn et al. [93, 94] made direct measurements of ions. These average values were used by Breneman and Stone [7] to find the power-law dependence of enhancements on *A/Q* in gradual SEP events. These ionization states, e.g. $Q_{Fe}$ = 14.1 ± 0.2 in gradual events, suggested source plasma temperatures of ~2 MK. Leske et al. [95] measured charge states at 15 – 70 MeV amu$^{-1}$ and Tylka et al. [96] at 200-600 MeV amu$^{-1}$. However, in impulsive SEP events they found $Q_{Fe}$ = 20.5 ± 1.2, and all elements up to Si were fully ionized. How could there be enhancements in Ne/O, Mg/O and Si/O by factors of ~2.5 [27] if all those elements had the same *A/Q* = 2.0? Subsequent measurements showed that observed ionization of Fe in impulsive SEPs varied with energy, as would be expected if the ions were stripped of some electrons after acceleration by passage through a small amount of material, suggesting acceleration at ~1.5 solar radii [97]. In contrast, measurements of velocity dispersion [98, 55, 56] and radio-emission onsets [64] in *gradual* SEP events suggest they begin at 2 – 3 solar radii, where densities are sufficiently low to preserve the charge states of the seed ions. However, it was clearly impossible to determine, from *direct* charge measurements, the *Q*-values those ions had during acceleration in impulsive events.

Hints or limits to charge states in impulsive sources could be found from the existence of enhancements of Ne, Mg, and Si, relative to C, N, O, and $^4$He. O is likely to be almost fully ionized but Ne could not be [27]. Later, average abundance enhancements in impulsive SEP events seemed to follow a power law in *A/Q* with *Q*-values appropriate for ~3 MK, both above 1 MeV amu$^{-1}$ [30] and even below [29]. How can we study variations in individual events?

After studying abundances in average impulsive SEP events [32], Reames et al. [33] made the brash assumption that the abundance enhancements *must* follow a power law. Of course, scattering theory does not produce perfect power-laws (see Appendix 1 of [99]), but they are usually a very good approximation. What remained was to fit the enhancements in an event at many assumed temperatures, using temperature to determine *Q* for each element via Mazzotta et al [100] for atomic numbers *Z* ≤ 28 and Post et al. [101] at higher *Z*, then choose the best fit from the minimum value of $\chi^2/m$ (where *m* is the number of degrees of freedom). An example of this process is shown in Fig. 2 for an impulsive event on 23 July 2016.

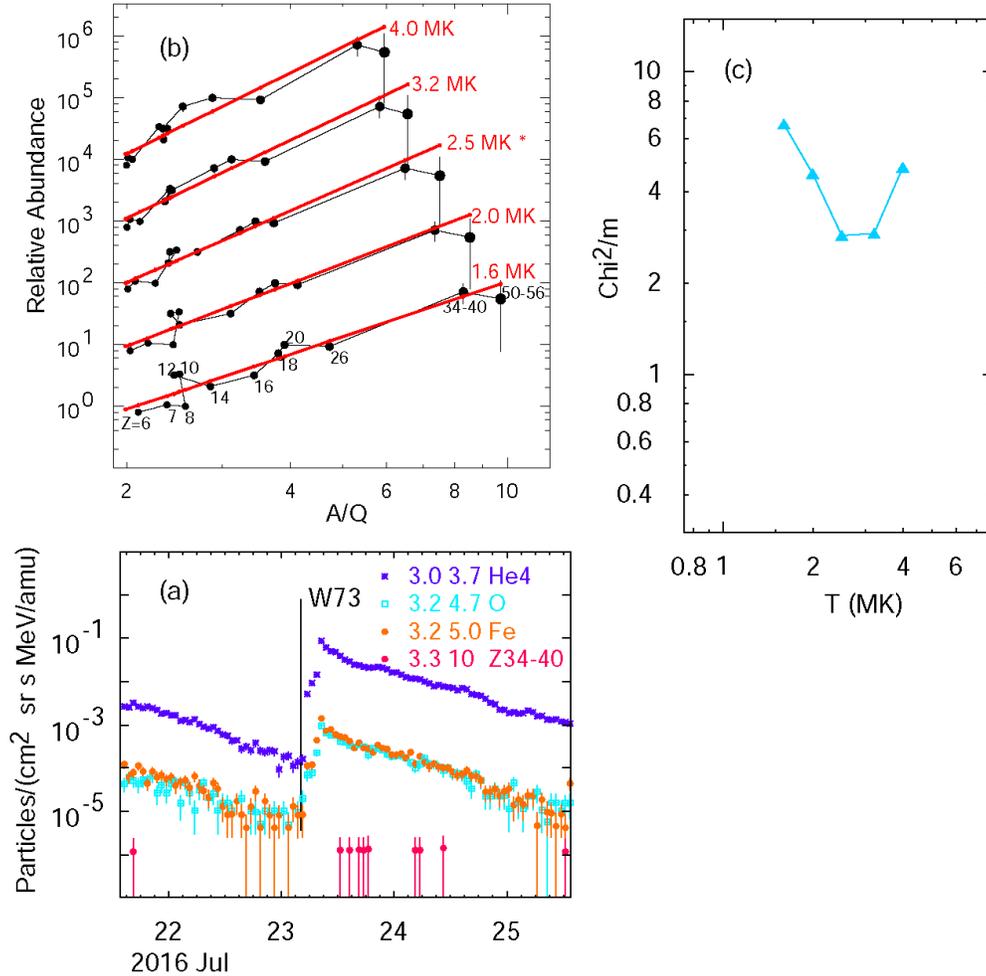

**Fig. 2** (**a**) Shows intensities of the listed ions and MeV amu$^{-1}$ intervals for an impulsive SEP event on 23 July 2016. (**b**) Abundance enhancements for ions with atomic number $Z \geq 6$, relative to O, are plotted and fitted vs. $A/Q$ at the listed temperatures (each shifted by a decade). $Z$ values are listed along the 1.6 MK fit. (**c**) Shows values of $\chi^2/m$ corresponding to each fitted temperature, with 2.5 MK barely selected as minimum for this event.

Surprisingly, the temperatures varied little from event to event. Of 111 impulsive events, Reames et al. [33] found 79 with minima at 2.5 MK and 29 at 3.2 MK and the powers varied from 2 to 8. Studies with variations in the weighting of the fits [102] made little difference. More-recent studies of EUV temperatures in the jet sources of $^3$He-rich events have confirmed a temperature of ~2.5 MK in these sources [103] – the SEP power-law fits and EUV measurements agree.

In all of these fits, a Maxwellian electron distribution was used to determine average $Q$ values vs. temperature. Recently Lee et al. [104] studied $A/Q$ dependence in impulsive SEP events using more-accurate kappa distributions that include high-energy tails on the electron distributions. They found that the source temperatures were not significantly changed even though a few $A/Q$ values varied at most by 10 – 30%. They noted that the SEPs must escape before any heating occurred. Lee et al. [104] is a very important study that establishes the validity of the technique and confirms the temperatures found in previous Maxwellian studies. Fits in the current article continue to use charges based upon Maxwellian distributions [100, 101] for simplicity and consistency with previous studies.

The differences in classes SEP1 through SEP4 begin to emerge when we compare proton abundances with the higher-$Z$ fits. If all of the particles come from a single population, it might be reasonable to include H and $^4$He in the fits, but when we begin to encounter events with two seed populations this no longer works. Examples of analysis of abundances in several impulsive SEP events are shown in Fig. 3.

The first two events in Fig. 3c (events 3 and 4) are clearly SEP1 events, since observed proton abundances agree well with the extrapolated fit of $Z \geq 6$ ions – strongly suggesting a single seed population from magnetic reconnection in the source jet. Event 5 in Fig. 3c could suggest that the enhancements begin above $^4$He, rather than H. This is a possible case according to theory of Drake et al. [44], but with a fast CME speed of 777 km s$^{-1}$ this event is very likely a SEP2 event.

Figure 3f shows the abundance pattern of a SEP2 or SEP3 event with a strong excess of H and $^4$He suggesting two seed populations for a shock provided by the 925 km s$^{-1}$ associated CME [32]. $^4$He is enhanced by having two seed contributions. This is probably a SEP3 event because of a similar preceding event and small fluctuations of the abundances – low $\chi^2/m$ – indicating averaging over multiple impulsive sources. However, we will see that most SEP3 events are gradual events lasting days while this event looks impulsive. The attempt to distinguish locally-sourced SEP2 events from broadly-sourced SEP3 events is sometimes difficult, but SEP3 events are found to have measurably smaller variations from averaging over many seed events (see Fig. 8 in [16]). Also, the separation of SEP1 and SEP2 events using proton excess sometimes depends upon uncertain physics, as shown in Fig. 3c.

Most SEP4 events have abundance enhancements that are quite flat; this lack of rigidity dependence suggests that the particles have undergone very little scattering in transit. This makes it difficult to determine source temperatures since $A/Q$ assignments do not matter. However, the temperatures that are found in SEP4 events are usually in the range 0.8 – 1.8 MK; this seems to be typical of ambient coronal plasma and helps distinguish this seed population from impulsive seed ion, which are at ~2.5 – 3.2 MK. Two sample SEP4 events, shown in Fig. 4, show slow suppression of the high-$Z$ enhancements that are initially flat, even though one event is a GLE [65] where trapping might be expected. The CME speed of the event in Fig. 4a is 1466 km s$^{-1}$: that in Fig. 4d was unmeasured but estimated as ~1420 km s$^{-1}$ [65]. In his initial study of enhancements vs. $A/Q$ in gradual events, Reames [16] included $^4$He in the fits. The presence of two seed components of $^4$He in SEP3 events causes erroneous secondary minima in $\chi^2/m$ at high temperatures, depending upon the assumed value of coronal $^4$He/O; this error was subsequently recognized, discussed, and corrected [105, 106]. Other properties of SEP4 events will be considered in Sect.4.

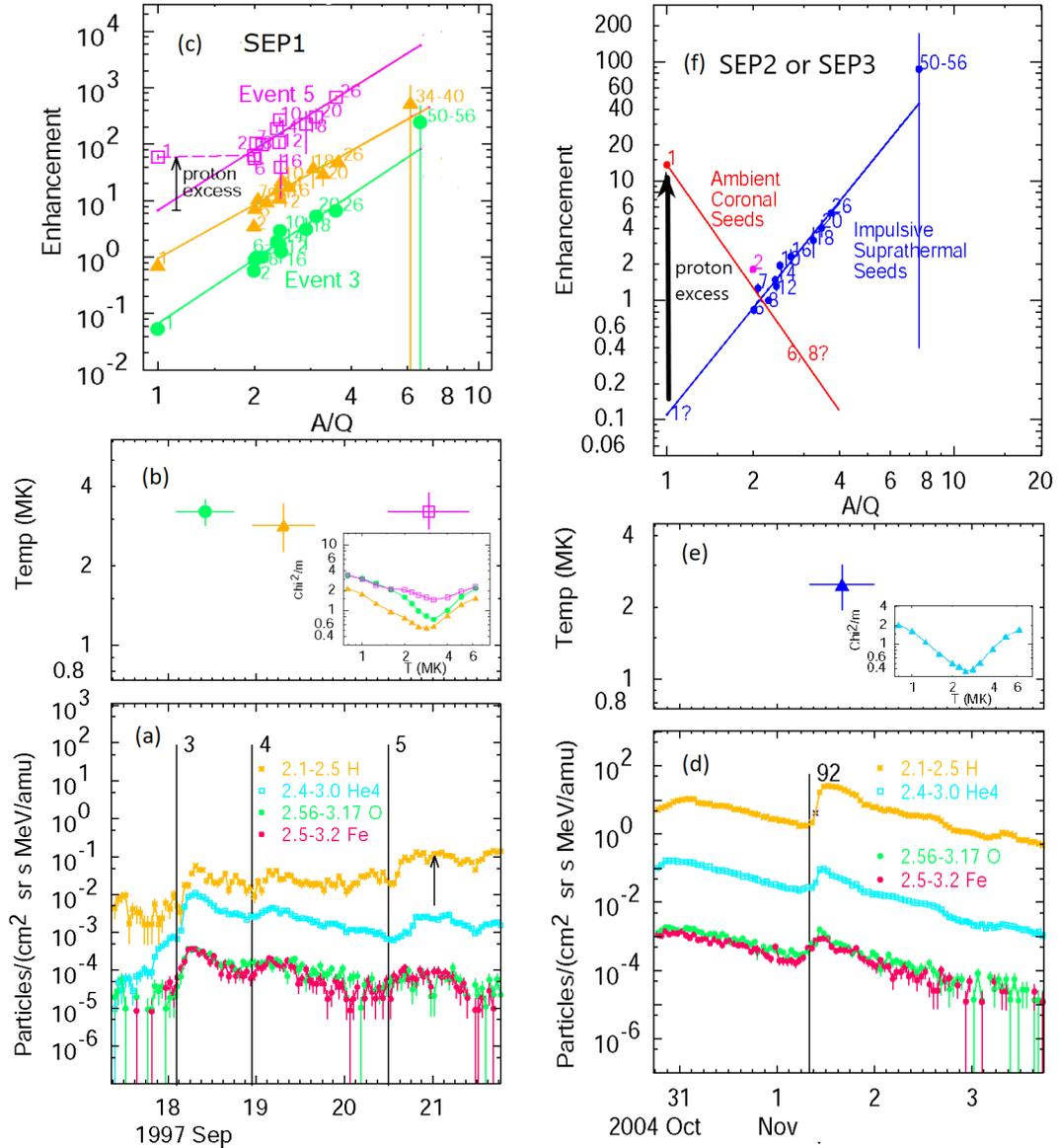

**Fig. 3** (**a**) and (**d**) show intensities of the listed ions and MeV amu$^{-1}$ intervals for impulsive SEP events. Panels (**b**) and (**e**) show the derived temperatures and event durations with insets showing minimization of $\chi^2/m$. In (**c**) and (**f**), abundance enhancements for ions, relative to O, are plotted and $Z \geq 6$ ions are fitted vs. $A/Q$ (each event shifted by a decade) with fits extended to protons at $A/Q = 1$. In (**c**), H and $^4$He agree with fits of other ions in Events 3 and 4 with a proton excess shown for Event 5; all are possible SEP1 events (see text). Panel (**f**) shows the interpretation of the evidence of two seed populations in a SEP2 or SEP3 event. Event numbers used refer to the event list in [32] (see Appendix).

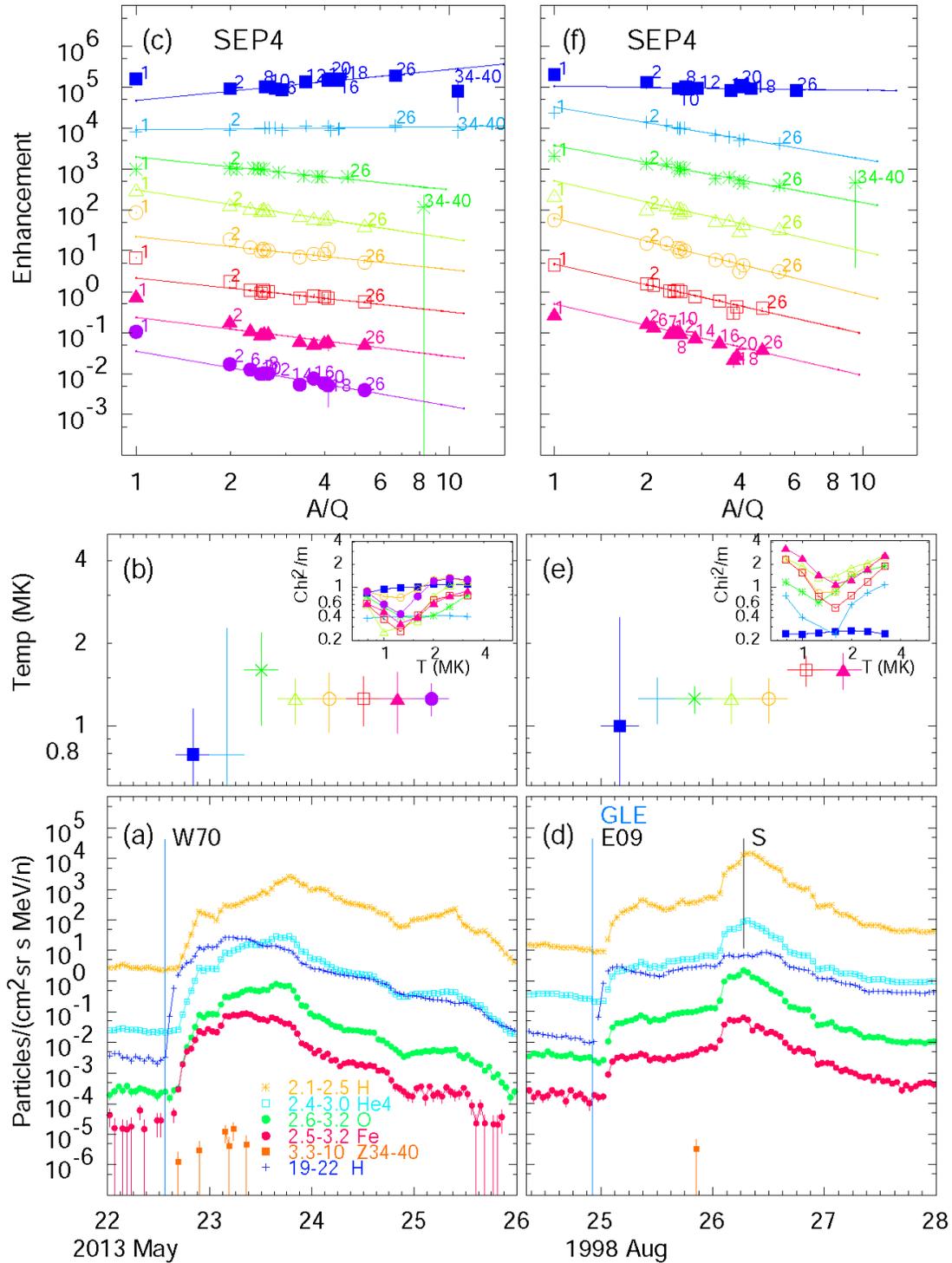

**Fig. 4** (**a**) and (**d**) show intensities of the listed ions and MeV amu$^{-1}$ intervals for two typical gradual SEP4 events. Panels (**b**) and (**e**) show the derived temperatures < 2 MK and time-interval durations with insets showing minimization of $\chi^2/m$. In (**c**) and (**f**), abundance enhancements for ions, relative to O, are plotted with colors and symbols of the time intervals below; sample values of $Z$ are listed, and $Z \geq 6$ ions are fitted vs. $A/Q$ (each interval shifted by a decade) with fits extended to protons at $A/Q = 1$. In SEP4 events, the fitted lines tend to intercept the measured protons. CME speeds are 1466 and ~1420 (see text) km s$^{-1}$, for the events in (**a**) and (**d**), respectively (see Appendix).

The fundamental logic used to distinguish the source populations and physical processes in this work is as follows [90]:

(1) The enhancements of the abundances of elements with $Z \geq 6$, relative to the average SEP coronal abundances, are assumed to be well represented as a power law function in $A/Q$. Best fits select the best temperature and corresponding $Q$-values for each time interval or event.

(2) Impulsive SEP1 events are found to have steeply ascending power laws and source temperatures of ~2.5 – 3.2 MK. Protons often fit the power-law, but sometimes perhaps the enhancement can begin above $^4$He. Any associated CMEs should have speeds < 500 km s$^{-1}$ (i.e. no shock reacceleration).

(3) Impulsive SEP2 events have characteristic steeply rising power-laws and ~2.5 – 3.2 MK temperatures of SEP1 component plus evidence of a shock-accelerated ambient seen as a large proton (and $^4$He) excess (as in Fig 3f) and/or CME speeds > 500 km s$^{-1}$.

(4) Gradual SEP4 events we find temperatures of ~0.8 – 1.8 MK (when power law is not too flat) and protons and $^4$He fit on (or occasionally below, see Sect 4.1) the high-$Z$ power law. Most events have flat or declining powers. In intense SEP4 events, wave amplification causes rising powers, with no proton excess. Powers suddenly decline after shock passage. SEP4 events are accompanied by fast, wide CMEs.

(5) Gradual SEP3 events have rising power laws and temperatures of ~2.5 – 3.2 MK (signature of SEP1 seeds) and a large proton (and $^4$He) excess (ambient coronal seeds, as in Fig. 3f). Abundances remain constant throughout the event. SEP3 events are accompanied by fast, wide CMEs.

In SEP1, SEP2, and SEP3 events, the source temperatures reflect the time when the enhancements were generated, e.g. 2.5 MK and $Q_{Fe} \approx 14$. If the charge states of these ions are measured later, i.e. at 1 AU, they will usually reflect the more highly ionized states of stripping [96] after acceleration, e.g. $Q_{Fe} \approx 20$, as found by Luhn et al. [93]. SEP4 ions will have ionization states typical of the ~1 MK corona, e.g. $Q_{Fe} \approx 10 - 13$.

## 3 Sequences of SEP3 Events

When we begin to scan solar cycle 23 for sequences of SEP3 events, probably fed by streams of impulsive events like that shown in Fig. 1, they begin early. Figure 5 shows a pair of SEP3 events at W33 then W63 in NOAA active-region 8100 as it rotates across the Sun. These are not small, obscure events, one a GLE. Ionization states and temperatures in SEP3 events are derived from the element enhancement pattern which was produced when the SEP1 seed particles were accelerated. As a possible source of seed particles, there is a flurry of Hα solar activity in the hours before the 4 November event, but not much prior to the event on 6 November. However, we would not expect to see both a SEP3 event and its seed population at 1 AU.

A second sequence of SEP3 events in May 1998 is shown in Fig. 6 with two GLEs from NOAA active-region 8210 as it rotates from W15 to W64. A smaller event partially obscured by a shock peak, occurs at W34 early on May 4; it can only be seen by an increase in Fe in Fig. 6a and an increase in derived temperatures in Fig. 6b, but any increases in elements below Fe are uncertain. There is moderate solar activity before each of the events of 2 May and 6 May, but it is not intense or unusual.

For completeness we also include in Fig. 7, the SEP3 event sequence recently presented in [107]. An early SEP2 event is followed by two SEP3 events that are both GLEs. These events occur in NOAA active region 9415 as it rotates off the west limb of the Sun in April 2001. This region shows no unusual activity prior to the SEP3 events. It seems that it would be very difficult to forecast the presence of seed particles for SEP3 events. In Fig. 7e the distribution seems to flatten with time for both protons and high $Z$, probably from a reduction in the impulsively-seeded component.

The events in these three sequences of SEP3 events are neither small nor obscure; they include five GLEs. GLEs are unrelated to abundances and the presence or absence of seed populations in any known way, but they are certainly evidence of energetic protons. Here they are a proxy for large intense gradual SEP events. However, there are also many SEP4 GLE events in active-region sequences as we shall see in the Sect. 4.

One of the important ideas for the existence of reaccelerated impulsive seed ions was that these faster ions could more-easily overtake the shock from downstream in quasi-perpendicular shocks where the angle between the direction of the field $B$ and the shock normal, $\theta_{Bn}$, is ~90° [81]. This may still be an important factor in critical cases, and may produce Fe-rich enhancements above ~30 MeV amu$^{-1}$ [81], but all of the SEP3 events in Figs. 5, 6, and 7 were not produced solely by quasi-perpendicular shock waves. These active regions must be sufficiently full of impulsive suprathermal seed ions that are readily accessible, perhaps like that suggested by Fig. 1, or perhaps some are even weakly bound in closed magnetic loops above the active region.

Note that the source plasma temperatures of 2 – 4 MK (nominally, $Q_{Fe} \approx 14$) from the abundance fits in Figs. 5b, 6b, 7b, and 8b, reflect the time when the abundance enhancements were generated, i.e. during the acceleration of the impulsive seed ions. After acceleration these impulsive ions are typically stripped [96] so they become more highly ionized (e.g. $Q_{Fe} \approx 20$ and $A/Q = 2$ for C – S [93]) before reacceleration by the shock in SEP2 or SEP3 events (e.g. [82]). Charge states of Fe measured in GLE events were tabulated by Mewaldt et al. [70]. The five GLEs in Figs. 5, 6, and 7 were the only ones with $Q_{Fe} \approx 20$, which is perfectly appropriate for SEP3 events where the impulsive seed ions are stripped to $Q_{Fe} \approx 20$ long before they encountered the shock acceleration that generated the SEP3 event. $Q_{Fe} \approx 20$ is a signature of fully-processed impulsive seeds (since 1987 [93]) and is unrelated to the matter density at SEP3 shock acceleration as Mewaldt et al. [70] suggest. We should also note, however, that direct measurement of a distribution of $Q_{Fe}$ could be better able to measure mixtures of seed populations, whereas our abundance-pattern technique determines only the dominant seed population at high $Z$.

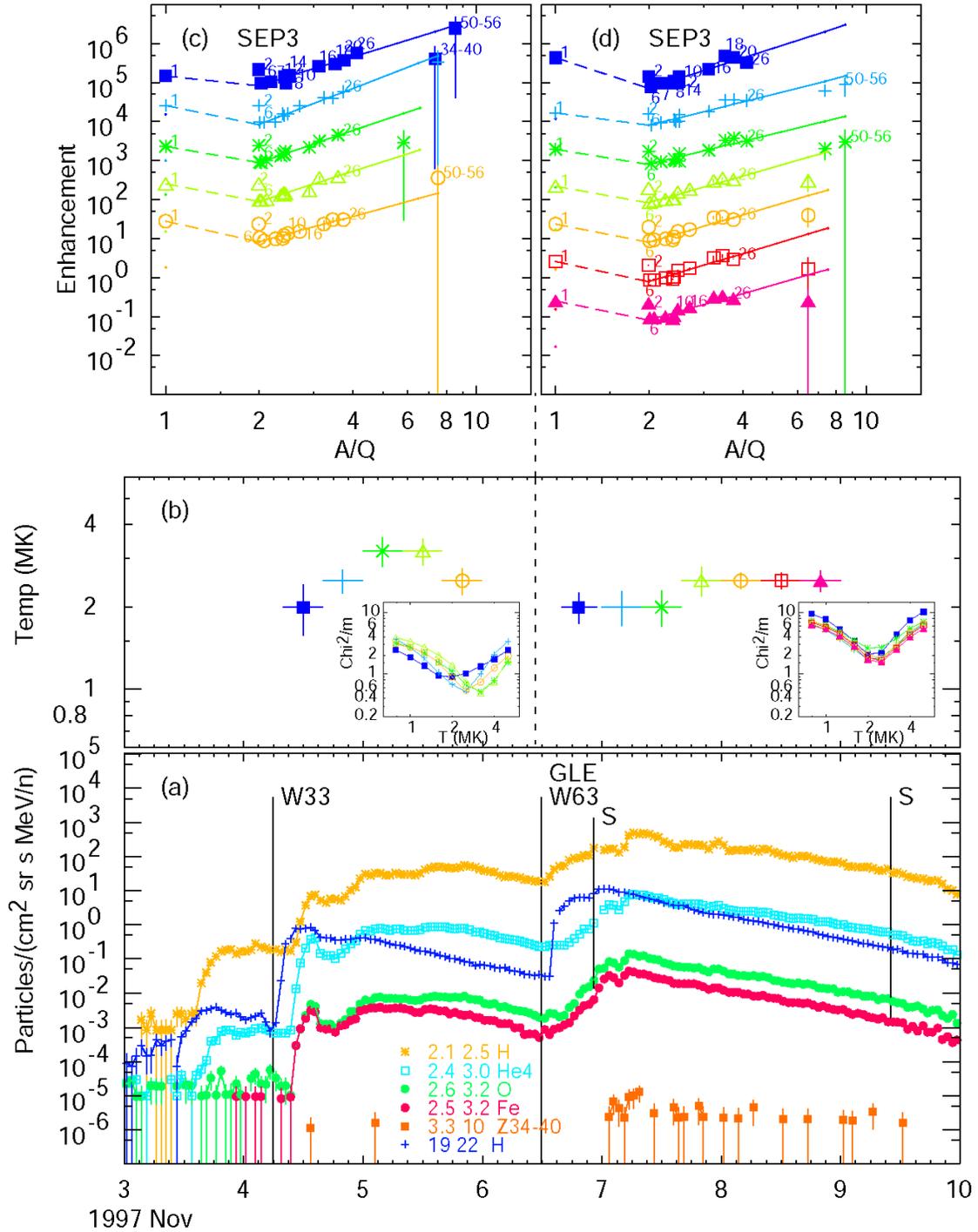

**Fig. 5** (**a**) shows intensities of the listed ions and MeV amu$^{-1}$ intervals during two gradual SEP3 events. Panel (**b**) shows the derived temperatures, all ≥ 2 MK, and time-interval durations with insets showing minimization of $\chi^2/m$. In (**c**) and (**d**), abundance enhancements for ions, relative to O, are plotted with colors and symbols of the time intervals below; sample values of Z are listed, and Z ≥ 6 ions are fitted vs. $A/Q$ (each interval shifted by a decade) with dashed lines drawn to the excess protons. CME speeds for the two events at W33 and W63 are 785 and 1556 km s$^{-1}$, respectively (see Table 1 in Appendix).

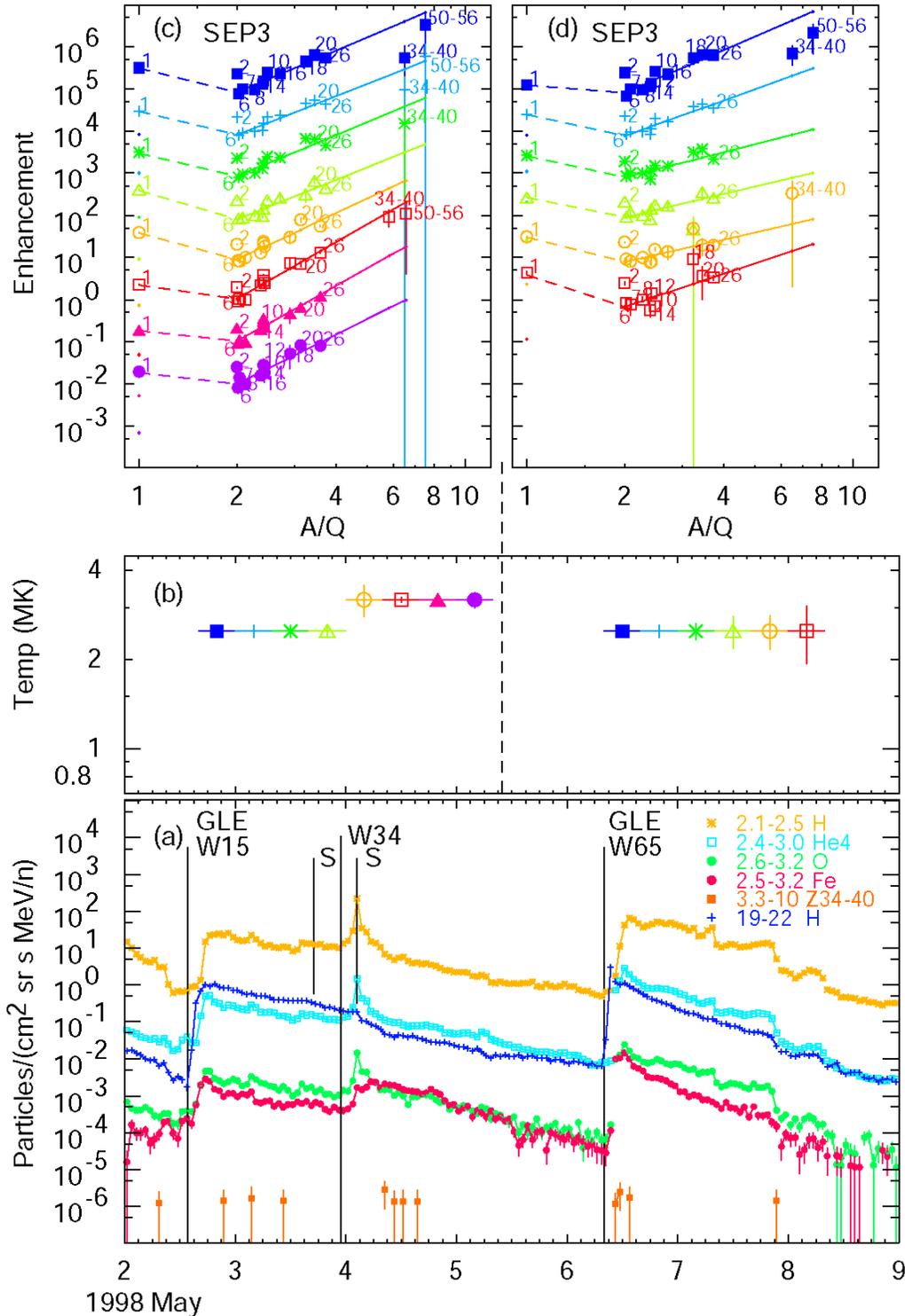

**Fig. 6**(**a**) shows intensities of the listed ions and MeV amu$^{-1}$ intervals during two gradual SEP3 events, both GLEs, with another injection on May 4 at W34 seen as increased Fe, Panel (**b**) shows the derived temperatures ≥ 2 MK and time-interval durations. In (**c**) and (**d**), abundance enhancements for ions, relative to O, are plotted with colors and symbols of the time intervals below; sample values of $Z$ are listed, and $Z \geq 6$ ions are fitted vs. $A/Q$ (each interval shifted by a decade) with dashed lines drawn to the excess protons. CME speeds of the events at W14, W34, and W65 are 938, 649, and 1099 km s$^{-1}$, respectively (see Table 1 in Appendix).

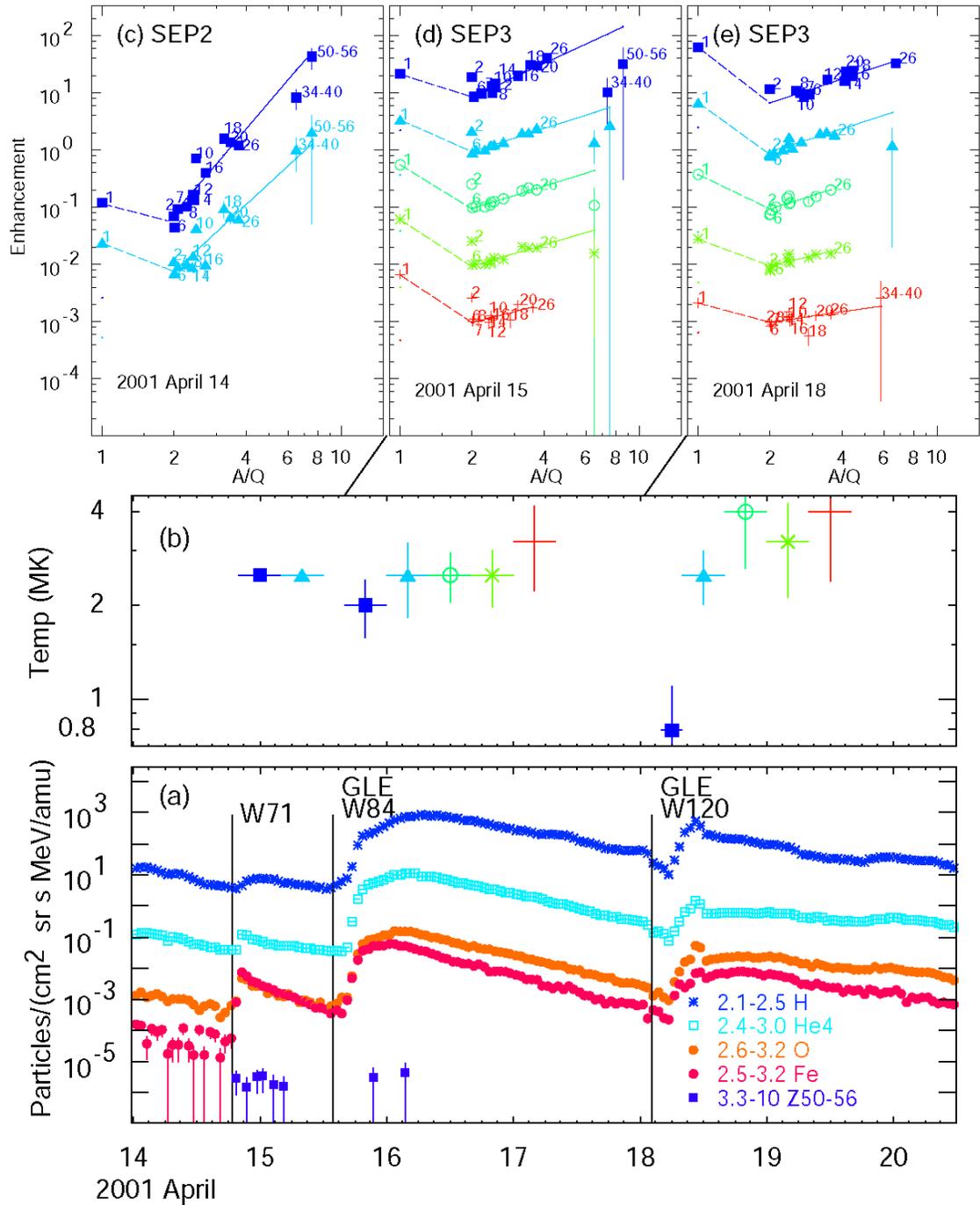

**Fig. 7** (**a**) shows intensities of the listed ions and MeV amu$^{-1}$ intervals during an impulsive SEP2 event and two gradual SEP3 events, both GLEs. Panel (**b**) shows the derived temperatures ≥ 2 MK and time-interval durations. In (**c**), (**d**) and (**e**), abundance enhancements for ions, relative to O, are plotted with colors and symbols of the intervals below; sample values of Z are listed, and Z ≥ 6 ions are fitted vs. $A/Q$ (each interval shifted by a decade) with dashed lines drawn to the excess protons. CME speeds for the events at W71, W84, and W120 are 830, 1199, and 2465 km s$^{-1}$, respectively (see Table 1 in Appendix).

### *3.1 Other Solar Cycles*

The reduced magnetic activity in solar-cycle 24 reduced the number and size of large SEP events and thus, sequences of SEP3 events are rarer in solar-cycle 24. Nevertheless, there is a sequence of SEP3 events early in the cycle that is shown in Fig. 8. It begins with a large SEP3 event at W36 on 4 August 2011 with source temperatures > 2 MK shown in Fig. 8b and enhancement patterns with modest proton excesses continuing smoothly through the shock as shown in Fig 8c. There is no change in the pattern after shock passage.

The small event at W61 on 8 August in Fig. 8a is probably a SEP2 event with rather large spread of the enhancements around the fit lines in Fig. 8d, especially a depression of S and enhancement of Ar. Actually these same differences seem to continue, to a lesser degree, into the abundances of the SEP3 at W69 on 9 August, suggesting that the earlier SEP2 event has dominated the impulsive seed population, It is a relatively rare occurrence that we can follow details of individual high-$Z$ element abundance patterns from one event to the next.. This may mean that the impulsive seed population did not persist for 5 days, but was renewed by the single large SEP2 event on 8 August.

Reames [90, 107] discussed aspects the solar-cycle variations of SEP3 events, which bears mention here. In this paper we have shown six SEP3 events (one coming in Fig. 9) that are GLEs out of a total of 16 GLEs in solar cycle 23 [65]. There are only two GLEs in solar cycle 24, both SEP4 events. Cohen et al. [108] studied abundances of H, He, O, and Fe in events seen on two or three spacecraft around the Sun at STEREO and near Earth during solar cycle 24 and found 41 events. Only one of them (shown in Fig.8) is a SEP3 event. SEP3 events are more probable early in the cycle (see Fig. 13 in [15] and Fig. 12 in Reames [90]) and all of the sequences shown here (Figs. 5 – 8) occur early in cycles.

Solar-cycle 25 is a weak cycle like cycle 24 and offers few SEP3 sequences up through December of 2023. A sequence on 21 and 23 July 2016 is probably only a SEP2, SEP3 sequence. However, one especially active and complex sequence showing impulsive, short-duration SEP3 events involves eight CMEs in two days and is shown in Fig. 9. In Fig 9a the CME onset times are flagged with CME speeds and the event intervals, numbered 149 through 154, are chosen to measure SEP peaks that are especially clear in O and Fe. In most cases these event intervals follow a CME onset by ~2 hours; events 150 and 152 are preceded by increases in 20 MeV protons. Fits to the abundance enhancements are shown in Fig. 9c with temperatures of 2.5 – 3.2 MK shown in Fig. 9b. These events show varying powers of $A/Q$ and proton excesses with two of the faster CMEs, 1815 and 1970 km s$^{-1}$ producing the smallest powers of $A/Q$ (2.7 and 3.2) in events 150 and 152, respectively, but no SEPs rise above background from the 1829 km s$^{-1}$ CME at 1341 UT on 16 July. Fit parameters are shown in Table 1. Can we be certain that all the events in Fig. 9 are SEP3s? There is always some doubt of the origin seeds, but here must be some sharing in such a complex active region. While the sequence in Fig. 9 is interesting, it is not a single pure SEP3 sequence as the others, since the three fastest CMEs probably come from behind the limb.

If we look backward in time, there are ~2 – 10 MeV amu$^{-1}$ abundance measurements from 1978 – 1984 during solar-cycle 21 by the Very-Low-Energy Telescope (VLET) on the ISEE-3 spacecraft [90, 27]. Of the seven GLEs measured by VLET, only one is a SEP3 event (10 April 1981), but there is no SEP3-event sequence associated with it.

<mark>Seeds and Sequences of SEP Events</mark> <mark>D. V. Reames</mark>

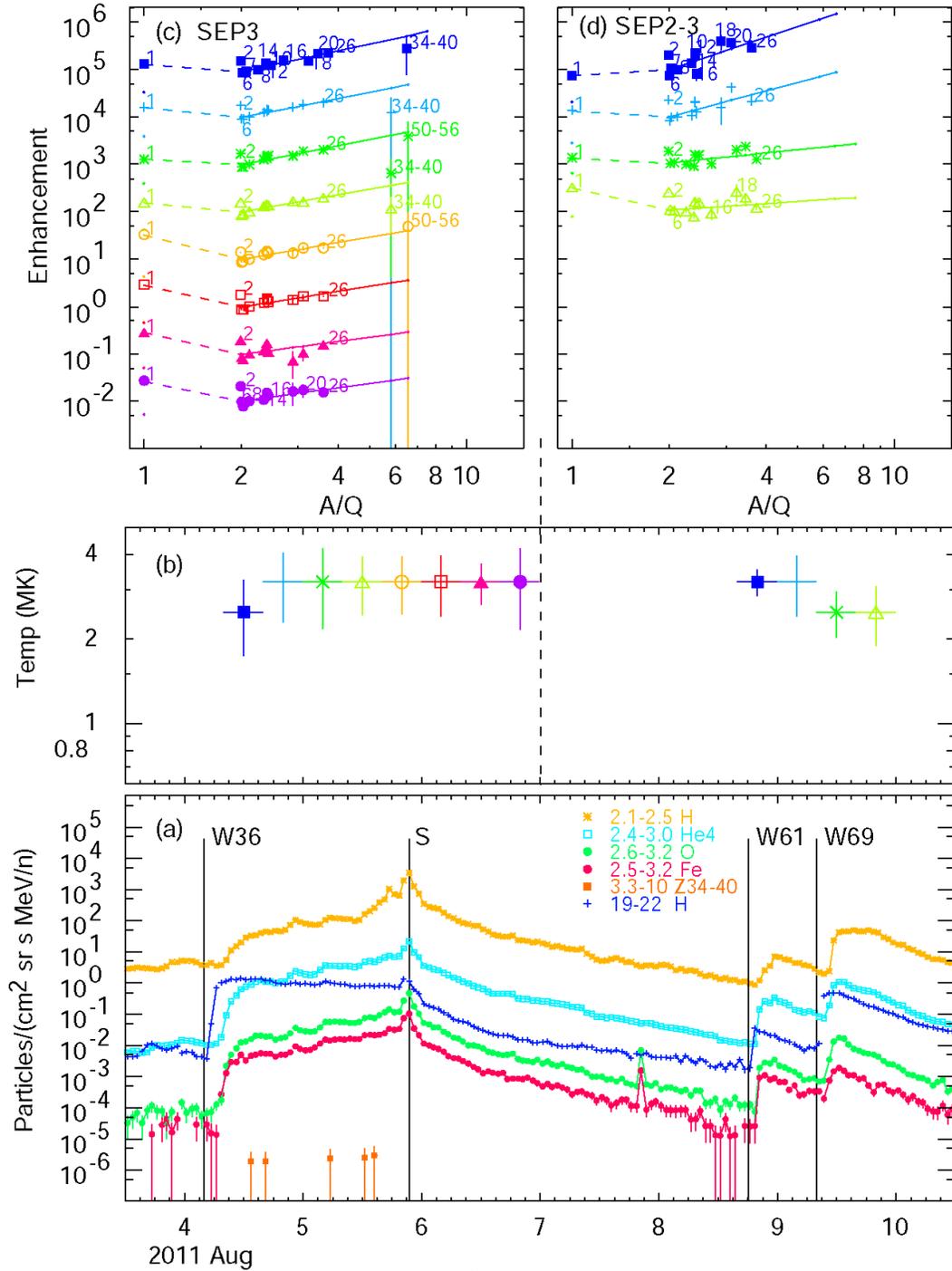

**Fig. 8**(**a**) shows intensities of the listed ions and MeV amu$^{-1}$ intervals during two gradual SEP events in solar-cycle 24 with a large SEP3 event at W36 followed later by a SEP2 event at W61 and a SEP3 event at W69. Panel (**b**) shows the derived temperatures ≥ 2 MK and time-interval durations. In (**c**) and (**d**), abundance enhancements for ions, relative to O, are plotted with colors and symbols of the time intervals below; sample values of Z are listed, and Z ≥ 6 ions are fitted vs. A/Q (each interval shifted by a decade) with dashed lines drawn to the excess protons. CME speeds are 1315, 1343, and 1610 km s$^{-1}$ (see Table 1 in Appendix).

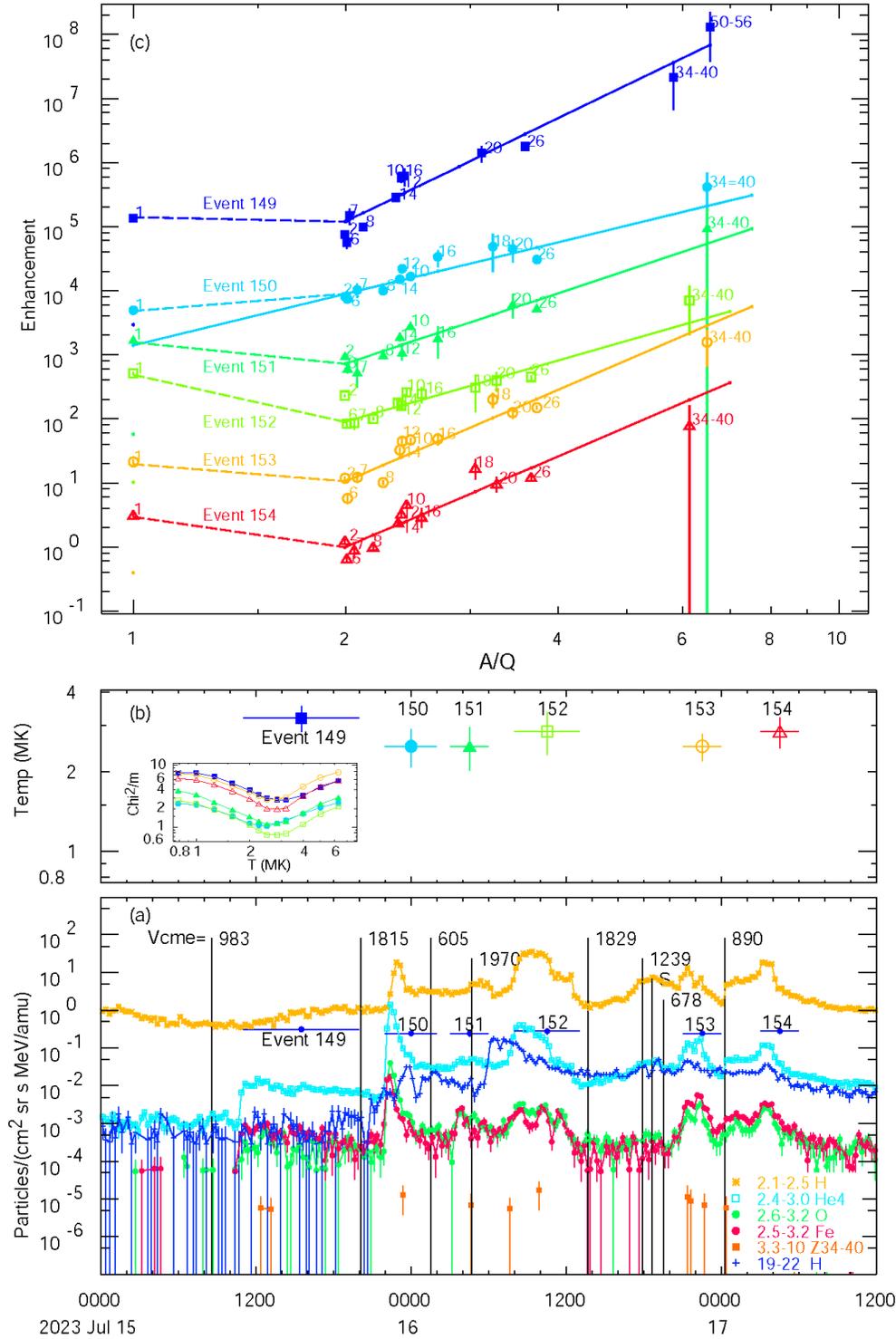

**Fig. 9**(**a**) shows intensities of the listed ions and MeV amu$^{-1}$ intervals during a series of SEP3 events in solar-cycle 25, flagged at their CME onset times with their speeds in km s$^{-1}$. Abundances are captured during event periods noted across the panel. Panel (**b**) shows the derived temperatures of the event intervals with an inset showing minimization of $\chi^2/m$. In (**c**) abundance enhancements for ions, relative to O, are plotted with colors and symbols of the time intervals below; sample values of $Z$ are listed, and $Z \geq 6$ ions are fitted vs. $A/Q$ (intervals shifted by a decades) with dashed lines drawn to the excess protons (see Table 1 in Appendix).

## 4 A Sequence with SEP4 Events

For SEP3 events, active regions are important as a likely source of the jets that feed the impulsive seed particles they need, but for SEP4 events, ambient coronal seed ions are everywhere. Nevertheless the evolution of an active region as it rotates across the Sun provides interesting new information for SEP4 events.  Figure 10 shows the evolution of NOAA active region 10486 which provided three GLEs in the Halloween series of events in October – November 2003. The basic features of element abundances in these events were first presented by Cohen et al. [109] who also included an earlier event from another active region.

The first event from E08 analyzed in Fig. 10c is typical for SEP4 GLEs [110, 107].  Self-amplified waves [111, 112] scatter all ions early [113 – 120], trapping them near the shock, especially H at low $A/Q$ =1, while heavier ions scatter less and leak out to 1 AU more easily.  All ion intensities rise to peak at the shock; behind the shock we pass into a reservoir – effectively a magnetic bottle with minimal scattering where intensities of trapped ions slowly decrease as the bottle expands [57].  Reservoirs were first seen as uniform intensities of ~20 MeV protons spread over ~180° in longitude  by McKibben [121], uniform intensities over 2.5 AU behind large events (Roelof et al. [122] who coined the name "reservoir"), and time-invariant spectral shapes late in large SEP events [61, 60].  Thus abundance enhancements in SEP4 GLEs rise with $A/Q$ before the time of shock passage and then suddenly decrease with $A/Q$ as seen in Fig. 10c, but the protons cling near the ends of the extended power-law fits as they vary.  Also the temperatures in Fig.10b for this event remain low, near ~1 MK.

Surprisingly, the second event from W02, analyzed in Fig. 10d, appears to be a SEP3 event. The analysis shows higher temperatures, ~2.5 MK, persistent heavy-element enhancements in Fig. 10d, and ten-fold proton excesses designated by the gray arrow.  Evidently active region 10486 has managed to generate a sufficient population of SEP1 ions on 29 October GLE that were not yet there for the previous event on 28 October.

After several days rest, region 10486 produces another GLE from W56 on 2 November.  This is clearly a SEP4 event with temperatures of ~1.3 MK and protons on the fit lines in the first and last intervals.  However, this event shows a new feature with protons *below* the fit lines at intermediate times as emphasized in the large gray circle in Fig. 10e. Similar proton suppression is circled in the next event from W83 in Fig. 10f.  This event is smaller and has power-law fits that are flat (rigidity independent) or slightly declining, making temperatures very hard to determine.

Before discussing the physics of proton-suppressed abundances we show in Fig. 11 two more events that soon followed those in Fig. 10.  The event on 2 December 2003 in Fig. 11a may be from the return of region 10486 on the next solar rotation and even the next gradual event on 11 April 2004 also seems related.  These two events and the two in November 2003 are the only four events out of ~100 in two solar cycles that show significantly "broken-down" protons which we discuss in the next section.  A few other events show hints of broken protons, but these four sequential events are the clearest by far.

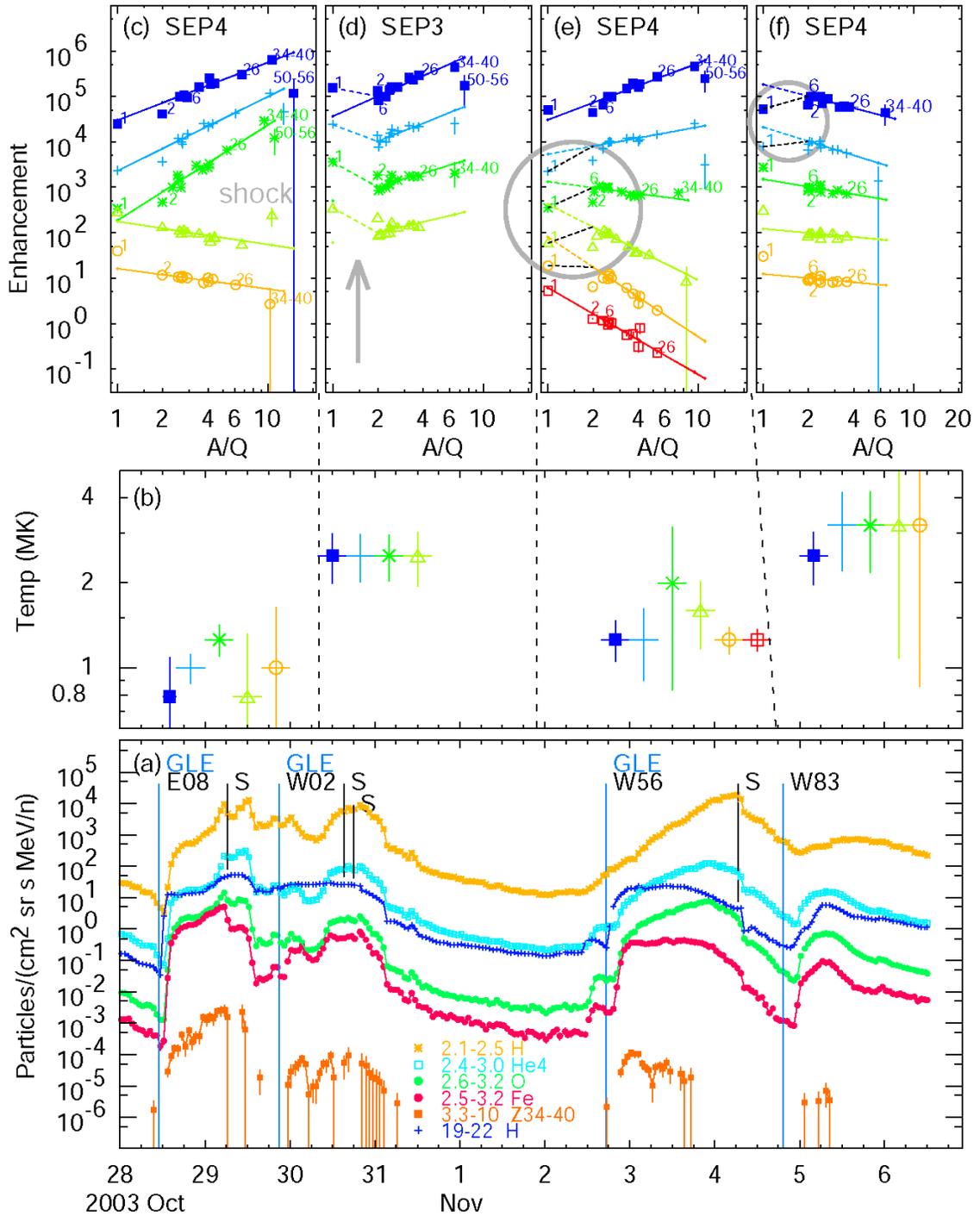

**Fig. 10** (**a**) shows intensities of the listed ions and MeV amu$^{-1}$ intervals during a series of four gradual SEP events, three of which are GLEs. Panel (**b**) shows the derived temperatures and time-interval durations. In (**c**), (**d**), (**e**) and (**f**), abundance enhancements for ions, relative to O, are plotted with colors and symbols of the time intervals below them; sample values of $Z$ are listed, and the $Z \geq 6$ ions fitted vs. $A/Q$ (each interval shifted by a decade) with dashed lines drawn to the protons in some cases. CME speeds are 2459, 2029, 2598, and 2657 km s$^{-1}$, respectively (see Table 1 in Appendix). Gray circles in (**e**) and (**f**) emphasize protons below the fit lines (see text).

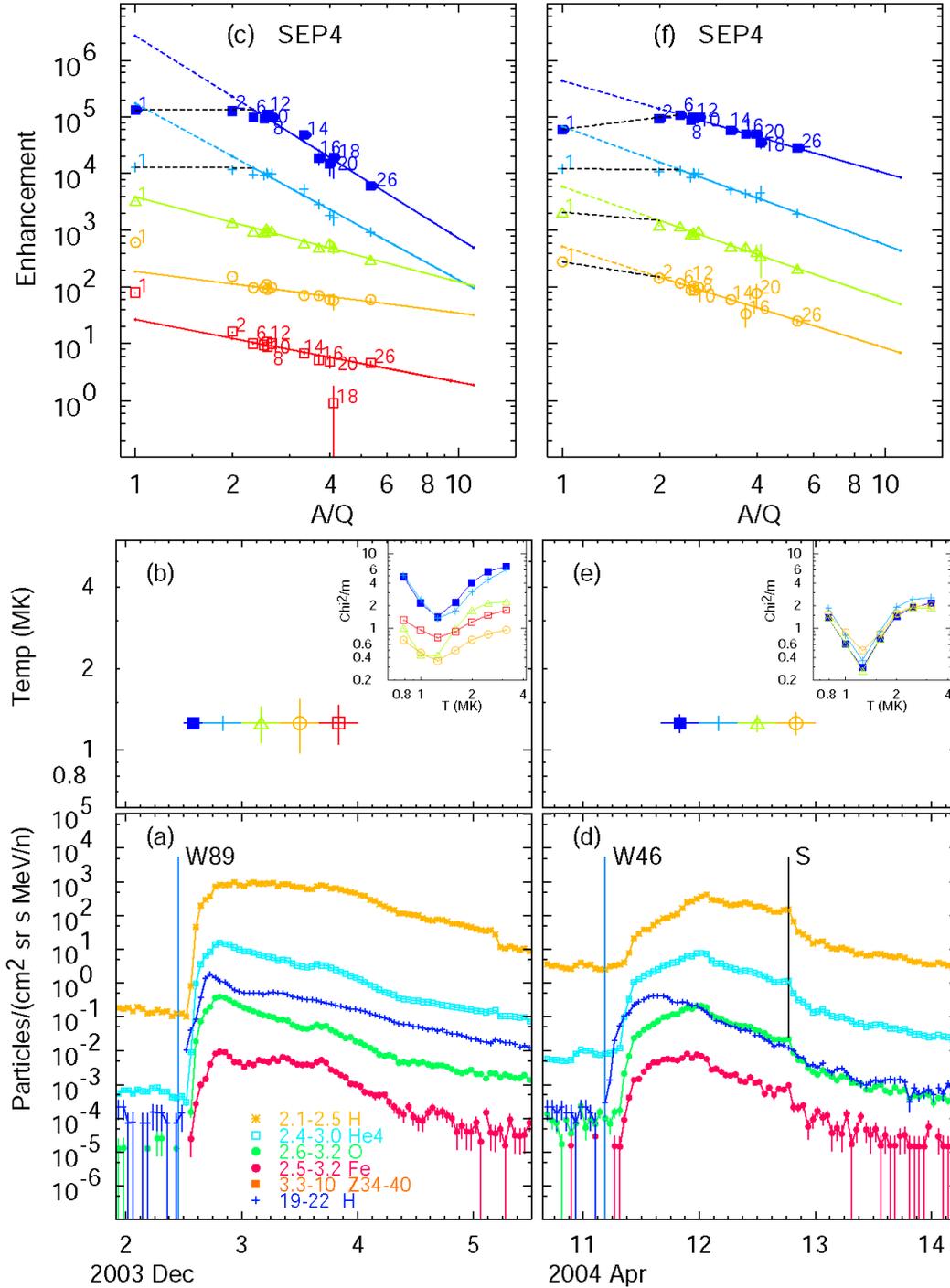

**Fig. 11** (**a**) and (**d**) show intensities of the listed ions and MeV amu$^{-1}$ intervals during two "broken proton" SEP4 events that follow those in Fig. 10. Panels (**b**) and (**e**) show the derived temperatures < 2 MK and time-interval durations with insets showing minimization of $\chi^2/m$. In (**c**) and (**f**), abundance enhancements for ions, relative to O, are plotted with colors and symbols of the time intervals below; sample values of $Z$ are listed, and $Z \geq 6$ ions are fitted vs. $A/Q$ (each interval shifted by a decade) with fits extended to protons at $A/Q = 1$. In these relatively rare SEP4 events, the protons fall below the measured fitted lines. CME speeds are 1393 and 1645 km s$^{-1}$ for events in (**a**) and (**d**), respectively (see Table 1 in Appendix).

Seeds and Sequences of SEP Events                                       D. V. Reames### *4.1 "Broken-Down" Proton Events*

If SEP ions scatter against magnetic turbulence or Alfvén waves, that scattering must depend upon the magnetic rigidity of the ion, hence upon *A/Q* for ions of the same velocity or energy-per-nucleon. In the events we have discussed, the rigidity of Fe is ~4 – 6 times that of protons. Our Fe scatters like ~100 MeV protons.

Consider the ion intensities in the GLE event of 2 November 2003 in Fig. 10a. Here, the protons rise continuously until the time of the shock passage, $^4$He and O rise to peaks early on 4 November, and Fe undergoes a flat peak on 3 November. We can identify three phases in these events: (1) Early in the event all species are trapped near the shock by the proton-amplified Alfvén waves in this large event [114, 115, 118 – 120], high-rigidity ions leak more easily so the intensities increase with *A/Q* as seen in the filled blue squares in Fig. 10e. (2) At intermediate times high-*A/Q* ions stop rising and begin to leak away, decreasing their intensities relatively, but protons are still trapped near the shock and on the rising part of the curve. (3) Finally, after crossing the shock all ions have declined; high-rigidity ions the most.

The other three events in Figs. 10f and 11 seem to skip the first phase. This is mainly because protons streaming from these smaller events are not able to cause sufficient wave growth to trap high-rigidity ions near the shock. With reduced trapping for a shorter period, the ions peak early and there is no rising *A/Q* dependence.

Larger GLEs like the one in Fig. 10c skip the second phase. Since all ions peak at the shock in these events there are no intermediate peaks; they all go from complete trapping of all species until they enter the reservoir behind the shock.

Reservoirs are especially important in these large events. While they have been reviewed elsewhere [57], the importance of self-amplified waves should be noted. Streaming protons amplify Alfvén waves that strongly limit ion intensities and flatten energy spectra on the plateau region early in events [119, 120, 115] where the escape of ions is controlled by the "streaming limit." Here the scattering of ions is strongly rigidity dependent. Viewed as a whole the shock structure and the upstream wave field traps protons much more effectively than Fe, at any given velocity. By the time this structure reaches 1 AU, much more of the Fe has already leaked away than have protons, so this "magnetic bottle" of the reservoir plus waves can contain a power-law distribution with depressed heavy-ion abundances like that of the red open squares in Fig. 10e. Note that wave-amplification greatly increases the effectiveness of "magnetic-bottle" reservoir formation in large events.

Why don't all SEP4 events have "broken" protons? With the exception of large GLE events, the answer seems to be that most SEP4 events involve minimal scattering during transport. Like the events in Fig. 4, they show flat (rigidity independent) or declining power laws in *A/Q*. The steepening decline with time measures the reduced scattering of higher-*Z* ions. Scattering must depend upon magnetic rigidity, hence upon *A/Q*; if there is minimal dependence upon *A/Q*, there must be minimal scattering.

The high temperatures in the last event in Fig. 10 suggest it could be a SEP3 event. The last three power-law fits in Fig. 10f, the protons do lie slightly (~2×) above the fit lines. We have chosen to assign the event a SEP4 because of the flat and declining fits and small proton excesses, attributing the high temperatures to the errors from flat fits. While it would be convenient to reserve temperatures above 2 MK for impulsive ions, we cannot exclude the possibility of ambient coronal ions from coronal regions above 2 MK, although it seems rare.

## 5 Discussion

Do most gradual SEP events involve minimal scattering during transport? Historically, it has been tempting to apply well-known diffusion theory to SEP profiles. Adjusting the diffusion coefficient arbitrarily, it is easy to fit the fast-rise slow-fall profiles of events from western sources, like that in Fig. 4a, even using a source delta function in time. The shock, centered 70° to the west, makes no obvious contribution to disrupt a nice diffusive picture. However, for the event in Fig. 4d, the shock peak seems to dominate, i.e. this source is certainly *not* a delta function, and hints that a similar shock source may secretively control the time variation of the event in Fig. 4a as well. It is easier to model diffusion than shock source evolution. Now, the small rigidity dependence of the power-law fits in most SEP4 events suggests minimal scattering. Once upon a time, the late slow decline of SEP intensities post-shock was also ascribed to slow diffusion rather than to an expanding reservoir; then sharp peaks of scatter-free $^3$He-rich events were seen to easily stream across the same reservoir that was supposedly highly turbulent and diffusive ([123] and Sect. 2.3.5 in [10]). Scattering and diffusion is often overestimated to explain SEP time profiles. Can our theories consistently explain the observed *A/Q* variation with time, and not just the proton time profile? The rigidity variation from *A/Q* is a new test of SEP transport theory. Scattering must depend upon magnetic rigidity.

Then with increasing intensities, the protons streaming away from the shock reach threshold intensities where wave-amplification becomes significant and scattering does become very important; ions are strongly trapped near the shock, acceleration is enhanced producing still-higher intensities, and the character of the event changes to that we see in many GLEs. Wave growth is essential in all shock events to scatter ions back and forth across the shock but in these large events it can dominate particle transport out to 1 AU as well. GLEs are a proxy for these intense events. Between them and the common smaller gradual SEP events with minimal scattering, lie the relatively rarer broken-proton events where different ions, with different scattering, peak at different times.

The events in Fig. 9 show some of the complexities in defining the source physics. These events might be called impulsive because of their abundances and durations, but they certainly differ from the shock-free impulsive events in Fig. 3a. Strong shocks dominate events in Fig. 9, and the impulsive abundances and durations seem almost coincidental. These strong shocks happen to occur in an active region that is rich with impulsive seeds. Some of the shocks are stronger in these impulsive SEP3 events than in the classical SEP4 gradual events in Fig. 4.

Do SEP3 events have the same type of variations that we see in SEP4 events? Yes, if we could normalize to the seed particle sources in SEP3 events rather than the coronal abundances, we might see the same effects. For example, the power-law slopes in *A/Q* do not change from positive to negative in SEP3 GLEs as they do in SEP4 GLEs, because SEP3s are dominated by the impulsive seeds throughout. The SEP3 GLE of 29

October 2003 (Fig. 10d) shows streaming-limit flattening of the low-energy spectra [119] but the SEP3 GLE of 2 May 1998 (Fig. 6a) does not because its proton spectrum has lower intensities at 10-100 MeV, i.e. too low to generate enough waves [119, 120, 115]. Any broken-proton periods may be lost in the excess protons of SEP3 events, hidden by their dual-seed character. However, most SEP3 events do not show significant differences in the peaking times of different components – suggesting minimal scattering. Also, they do not show a change in power-law slope when crossing the shock.

In principle, a large gradual SEP3 event can have heavy-element enhancements from both impulsive seed ions and wave generation by streaming protons, and we do find some evidence of such effects. The SEP3 event of 29 October 2003 (Fig. 10) does show suppressed early proton spectra below 10 MeV (see Fig. 3 in [119]), but the suppression is much less than that in the 28 October 2003 SEP4 event, especially at higher $Z$. Even among GLEs, the proton intensities at ~10 MeV (that could potentially amplify Alfvén waves) tend to be lower in the observed SEP3 events. Generally, time variations in observed SEP3 events do not suggest significant proton trapping or sudden abundance changes at the shock as seen in large SEP4 events.

Three of the four broken-proton events (November-December 2003) seem to emerge from a common active region and the fourth event (April 2004) soon follows. Could a single active region somehow cause the rigidity-dependent turbulence that could explain this sequence of different-sized events? Could it persist for months? This seems very unlikely, and no known mechanism comes to mind, but coincidence also seems unlikely.

The degree of accuracy and validity of the approximation of a power-law in rigidity can be judged from the many events in figures in this, and other, articles. However, we should mention that below 1 MeV amu$^{-1}$ there is evidence for resonant mechanisms that produce abundance peaks instead of power-laws. Mason et al. [124] found spikes in the abundances of Si or S in 16 events in 16 years, clearly from a resonant process that enhanced a limited region of $A/Q$. The second harmonic of the waves that enhance $^3$He at $A/Q = 1.5$ might enhance Si or S if $A/Q = 3$. Possible abundances accompanying $^3$He-rich events were studied by Roth and Temerin [125]. However, abundances in these few events that are S-rich at suprathermal energies < 1 MeV amu$^{-1}$ were examined with Wind/LEMT above 2 MeV amu$^{-1}$ and were found to be too small to be measured, despite LEMT's 50-fold larger geometry factor. This suggests that the resonant process produces very steep spectra. Recently, Mason et al. [126] found other events where higher harmonics of the waves that enhance $^3$He, subsequently controlled by wave damping by $^4$He ions with $A/Q = 2$, that could explain the greater enhancement of ions with $A > 100$. These events could not be observed near Earth. Unusually large variations of specific elements from the fit lines, especially in SEP1 and SEP2 events like those discussed above for Fig. 8d, could come from resonant enhancements, especially when suprathermal ions are reaccelerated by shocks. There have also been attempts to identify the origin and properties of occasional large enhancements of Ne [127].

It is certainly possible that there are small pockets of magnetic reconnection where unusual conditions, including temperatures, produce unusual abundance variations at suprathermal energies, including ions from resonances. However, the dominant behavior of SEPs from energetic solar jets, even below 2 MeV amu$^{-1}$, is surely still a power law in *A/Q*, on average. For individual impulsive SEP events, the power of *A/Q* is certainly not constant at 3.6, but varies from 2 to 8 [33]. We should also comment that abundances above ~10 MeV amu$^{-1}$ might be affected by source-spectral breaks which also depend upon *A/Q*, making the region 2 – 10 MeV amu$^{-1}$ the ideal interval for power-law abundance studies. Amplitudes of the fluctuations in enhancements in SEP3 events are observed to be even smaller since they average over several SEP1 or SEP2 events (e.g. Fig. 8 in [33]).

## 6 Summary

The assumption that element abundance enhancements or suppressions in SEP events depends upon a power law in ion magnetic rigidity or gyro-radius, and hence upon *A/Q*, has lead to best-fit determinations of SEP source plasma temperatures during initial acceleration. At energies 2 – 10 MeV amu$^{-1}$, impulsive SEP events from magnetic reconnection in solar jets have temperatures of ~2.5 MK (e.g. $Q_{Fe} \approx 14$), which differ from those in shock-accelerated ambient coronal plasma that are typically 0.8 – 1.8 MK. Note, however, that while abundances reflect temperatures and ionization states during acceleration, the charge states of impulsive ions and seeds observed later reflects stripping after acceleration (e.g. $Q_{Fe} \approx 20$), ambient coronal ions do not (e.g. $Q_{Fe} \approx 10 - 13$) Impulsive SEP events always have steeply enhanced abundances, rising, on average, as the 3.6 power of *A/Q* (2 – 8 power in individual events) while in most gradual SEP events CME-driven shock waves accelerate ambient ~1-MK coronal plasma to produce flat (rigidity independent) or declining power-law behavior. Lee et al. [104] have confirmed the temperature analysis in impulsive SEP events using more-accurate kappa distributions.

However, more-intense gradual SEP events can produce heavy-element enhancements in two different ways. (1) Accelerated intensities of energetic protons can become so high that the streaming protons amplify Alfvén waves that scatter subsequent ions, trapping them near the shock; protons are trapped very efficiently while higher-rigidity ions leak out, with all intensities rising until shock passage, then falling. (2) Shocks can reaccelerate impulsive ions (with their signature ~2.5 MK source temperatures later stripped to $Q_{Fe} \approx 20$) from a streaming seed population from profuse impulsive events from an active region (Fig. 1); these pre-enhanced abundances dominate over ions from ambient-coronal seeds in the SEP heavy element region, but cannot dominate ambient-coronal protons (and sometimes $^{4}$He). Measured $Q_{Fe} \approx 20$ values in large SEP3 GLEs are a property of the impulsive seed ions, not of the SEP3 shock environment.

Thus, the combination of two acceleration mechanisms and two seed populations for shocks has led to the suggestion of four possible classes of events: SEP1 impulsive events from "pure" reconnection in jets, SEP2 impulsive events with reacceleration by a local shock, SEP3 gradual events with heavy elements dominated by shock reacceleration of preexisting impulsive seeds, and SEP4 gradual shock-accelerated events dominated entirely by ambient-coronal seeds.

During solar-cycle 23, three different active regions continued producing enough impulsive seed particles, as they rotated across the Sun for a week, to allow repeated large, gradual SEP events to reaccelerate enough impulsive seed ions to dominate their heavy-element abundances.  Most of these SEP3 events are large GLEs and no SEP4 events occur in these three sequences.

      Earlier suggestions that impulsive seed particles are preferentially enhanced in quasi-perpendicular shocks may still be true, but these sequences with consecutive SEP3 events are unlikely to involve only quasi-perpendicular shocks.  Presumably the impulsive seed particles are sufficiently plentiful that they dominate SEP elements C and above.  Thus quasi-perpendicular shocks may be sufficient but not necessary condition when impulsive seed particles are plentiful.

      In contrast, another very active region produces a sequence SEP4, SEP3, SEP4, SEP4, showing that impulsive-seed-particle dominance can come and go.  At present it would be difficult to predict in advance that conditions exist for an impulsively-seeded SEP3 event, although a series of such events would certainly suggest the next one.

      The solar-cycle variation in SEP3 events is fairly dramatic.  Where solar-cycle 23 produced three week-long sequences of large SEP3 events (mostly GLEs), plus other individual SEP2 events, solar-cycle 24 produced only one modest sequence of SEP3 events.  SEP3 events and sequences tend to occur early in a cycle.  Perhaps the early cycle produces a few large active regions capable of both many jets and several large CMEs, while active regions in the late cycle are more fragmented.

      While high-intensity SEP4 events have previously been observed to amplify waves that trap ions, so as to produce enhanced power-law abundances, here we have observed rarer events where protons are well trapped but heavier ions leak more freely, with intensities peaking earlier at higher *A/Q*.  This produces power laws that rise from protons then descend above He.  These "broken-proton" events have not been reported previously.

      Since power-laws in abundance enhancements were first displayed by Breneman and Stone [7], they have become increasingly useful in the study of SEP events.  Previously, ratios such as Fe/C or Fe/O served as a proxy for power-law enhancements, but those ratios alone provide no ionization-state or temperature estimates. Also, the unique ~2.5 MK source temperature in impulsive events is soon erased by stripping after acceleration.  It is these stripped ions that can be reaccelerated by shocks.  In contrast, ambient coronal seed ions are found to have temperatures of ~1 – 1.8 MK, helping distinguish the two seed populations at high *Z*.  Subsequently, power-law fits gained new life by distinguishing events where the protons (and $^4$He) come from the same seed population as the heavier ions (i.e. fit the power law) or from a different one (exceed the power law).  Now, in SEP4 events, power-law fits provide a new measure of the rigidity dependence and strength of scattering and of wave growth, independently of the spectra or time variations of proton intensities.  Thus power laws in *A/Q* provide a powerful way to distinguish rigidity and velocity dependence in acceleration and transport – an increasingly useful tool.

**Disclosure of Potential Conflicts of Interest** The author declares he has no conflicts of interest.

# Appendix

Table 1 shows properties of the SEP events studied in Figs. 2 through 11. Columns show the figure number, the onset time of an event (usually the CME observed time, if any), identification number shown on the figure or GLE, the probable SEP type number, SEP1-4 [90], the first-order, plane-of-sky CME speed [128], the probable solar source location, and the best-fit source temperature and power of $A/Q$ derived from fitting the abundance enhancements vs. $A/Q$. For gradual events with multiple time intervals the last two columns are an average or, in two cases, typical pre- and post-shock values.

Table 1 Properties of SEP Events in Figures

| Fig. | Onset | ID on Fig. | SEP# | $V_{CME}$ Km/s | Location | Temp MK | Power of $A/Q$ |
|---|---|---|---|---|---|---|---|
| 2 | 2016 Jul 23 0524 | 121 | SEP2 | 835 | N05W73 | 2.8±0.3 | 4.0±0.4 |
| 3 | 1997 Sep 18 0215 | 3 | SEP1 | - | S25W76 | 3.2±0.5 | 3.8±0.4 |
| 3 | 1997 Sep 18 2020 | 4 | SEP1 | 343 | S18W90 | 2.8±0.5 | 3.1±0.3 |
| 3 | 1997 Sep 20 1020 | 5 | SEP2 | 777 | NA | 3.2±0.5 | 3.6±0.7 |
| 3 | 2004 Nov  1 0606 | 92 | SEP3 | 925 | N12W90 | 2.5±1.4 | 3.0±0.2 |
| 4 | 2013 May 22 1326 | - | SEP4 | 1466 | N15W70 | 1.2±0.4 | -0.8±0.5 |
| 4 | 1998 Aug 24 2212 | GLE | SEP4 | ~1420 | N35E09 | 1.3±0.5 | -1.4±0.4 |
| 5 | 1997 Nov  4 0610 | - | SEP3 | 785 | S14W33 | 3.2±0.6 | 2.5±0.6 |
| 5 | 1997 Nov  6 1210 | GLE | SEP3 | 1556 | S18W63 | 2.5±0.5 | 2.3±0.2 |
| 6 | 1998 May  2 1406 | GLE | SEP3 | 938 | S15W15 | 2.5±0.5 | 3.1±0.3 |
| 6 | 1998 May  3 2202 | - | SEP2 | 649 | S13W34 | 3.2±0.5 | 4.0±0.5 |
| 6 | 1998 May  6 0829 | GLE | SEP3 | 1099 | S11W65 | 2.5±0.5 | 2.0±0.5 |
| 7 | 2001 Apr 14 1754 | - | SEP2 | 830 | S18W71 | 2.5±0.6 | 4.3±0.6 |
| 7 | 2001 Apr 15 1306 | GLE | SEP3 | 1199 | S20W84 | 2.5±0.8 | 1.3±0.5 |
| 7 | 2001 Apr 18 0230 | GLE | SEP3 | 2465 | S?W120 | 3.0±1.0 | 1.4±0.5 |
| 8 | 2011 Aug  4 0412 | - | SEP3 | 1315 | N19W36 | 3.2±0.8 | 1.3±0.2 |
| 8 | 2011 Aug  8 1812 | - | SEP2 | 1343 | N16W61 | 3.2±0.5 | 2.0±0.5 |
| 8 | 2011 Aug  9 0812 | - | SEP3 | 1610 | N17W69 | 2.5±0.5 | 0.6±0.4 |
| 9 | 2023 Jul 15 0833 | 149 | SEP3 | 983 | S23W37 | 3.2±0.5 | 5.3±0.7 |
| 9 | 2023 Jul 15 2006 | 150 | SEP3 | 1815 | S22W45 | 2.5±0.5 | 2.7±0.4 |
| 9 | 2023 Jul 16 0133 | 151 | SEP3 | 605 | S20W35 | 2.5±0.5 | 3.7±0.5 |
| 9 | 2023 Jul 16 0442 | 152 | SEP3 | 1970 | S23W49 | 2.8±0.5 | 3.2±0.4 |
| 9 | 2023 Jul 16 1753 | 153 | SEP3 | 1239 | S23W58 | 2.5±0.5 | 4.7±0.6 |
| 9 | 2023 Jul 17 0016 | 154 | SEP3 | 890 | - | 2.8±0.4 | 4.7±0.6 |
| 10 | 2003 Oct 28 1110 | GLE | SEP4 | 2459 | S16E08 | 1.0±0.4 | 1.6 to -0.5 |
| 10 | 2003 Oct 29 2054 | GLE | SEP3 | 2029 | S15W02 | 2.5±0.5 | 1.3±0.5 |
| 10 | 2003 Nov  2 1730 | GLE | SEP4 | 2598 | S14W56 | 1.2±0.4 | 1.2 to -1.7 |
| 10 | 2003 Nov  4 1954 | - | SEP4 | 2657 | S19W83 | 3.0±1.0 | -0.8±0.5 |
| 11 | 2003 Dec  2 1050 | - | SEP4 | 1393 | S14W89 | 1.3±0.4 | -3.0±0.8 |
| 11 | 2004 Apr 11 0430 | - | SEP4 | 1645 | S16W46 | 1.3±0.4 | -1.9±0.5 |